# ON THE PERFORMANCE POTENTIAL OF CONNECTION FAULT-TOLERANT COMMIT PROCESSING IN MOBILE ENVIRONMENT

Tome Dimovski and Pece Mitrevski

Faculty of Technical Sciences, St. Clement Ohridski University, Bitola, Macedonia
{tome.dimovski, pece.mitrevski}@uklo.edu.mk

## ABSTRACT

*Mobile inventory, mobile commerce, banking and/or commercial applications are some distinctive examples that increasingly use distributed transactions. It is inevitably harder to design efficient commit protocols, due to some intrinsic mobile environment limitations. A handful of protocols for transaction processing have been offered, but the majority considers only a limited number of communication models. We introduce an improved Connection Fault-Tolerant model and evaluate its performance potential by comparing results in several deferent scenarios, as well as its contribution to the overall mobile transaction commit rate. Our performance analysis, conducted using general-purpose discrete-event simulation programming language, evaluates the effect of (i) ad-hoc communication between mobile hosts and (ii) the employment of an appropriate decision algorithm for mobile host agents. Conjointly, they substantially improve commit rate. In respect of the stochastic nature of random network disconnections, we determine connection timeout values that contribute the most to the highest perceived ad-hoc communication impact.*

## 1. INTRODUCTION

The increasing emergence of mobile devices contributes to a rapid progress in wireless technologies. Mobile devices interacting with fixed devices can support applications such as: e-mail, mobile commerce (m-commerce), mobile inventory, etc. Yet, there are many issues that are challenging and need to be resolved before enabling mobile devices to take part in distributed computing. For distributed systems, a transaction is a set of operations that fulfill the following condition: either all operations are permanently performed, or none of them are visible to other operations (known as the *atomicity* property). Therefore, the key issue in the execution of transactions is the protocol that ensures atomicity.

The mobile environment is comprised of mobile devices with limited resources, e.g. processing, storage, energy capacity and continuously varying properties of the wireless channel. Wireless communication induces much lower bandwidth, higher latency, error rates and much higher costs. This, in turn, increases the time needed for *Mobile hosts (MHs)* to execute transactions and can even lead to execution failure [1-4]. MHs are highly susceptible devices because they are easily damaged or lost and, in addition, they naturally show frequent and random network disconnections. These limitations and characteristics of the mobile environment make it harder to design appropriate and efficient commit protocols [5]. A protocol that aborts the transaction each time the MH disconnects from the network is not suitable for mobile environments, simply because it is considered to be a part of the normal mode of operation. In other words, disconnections need to be *tolerated* by the protocol. A solution for the management of *fault tolerance* in ad-hoc networks has been proposed in [6].

The *Two-Phase Commit (2PC) protocol* [7] that allows the involved parties to agree on a common decision to commit or abort the transaction even in the presence of failures is the most commonly used protocol for fixed networks but is *unsuitable* for mobile environments [8]. A handful of protocols for transaction execution in distributed mobile environment have been offered [9-19], but the majority considers only a limited number of communication models.

The main contribution of this paper is a simulation-based performance analysis of an amended *Connection Fault-Tolerant (CFT)* model for mobile distributed transaction processing, designed to show resilience to connection failures of mobile devices and initially introduced as a conceptual model in [20]. It differs from other infrastructure based protocols [9-11], [13], in the fact that, besides the standard communication between MHs and the fixed network, it (i) supports *ad-hoc communication* between MHs and (ii) introduces a *decision algorithm* that is responsible for decision making on behalf of a mobile host in a special case, when neither standard, nor ad-hoc communication is possible. We evaluate the performance potential of the CFT model by comparing the results in several deferent scenarios, as well as its contribution to the overall mobile transaction commit rate. In respect of the stochastic nature of random network disconnections, we determine connection timeout values that contribute the most to the highest perceived ad-hoc communication impact.

Section 2 provides a short survey of related work. In Sections 3 and 4 we formalize the model of the mobile environment and the transaction model, respectively. A comprehensive description of the Connection Fault-Tolerant commit processing framework is given in Section 5, whereas in Section 6 we present simulation results and their analysis. Section 7 discusses conclusions.

## 2. RELATED WORK

In order to reach a final transaction termination decision in any message oriented system, the *Transaction Commit on Timeout (TCOT)* protocol [9] can be universally used. Based on a timeout approach for Mobile Database Systems, this protocol limits the amount of communication between the participants in the execution. It decreases the number of wireless messages during execution, and does not consider MHs as active participants in the execution of transactions.

The basic idea of the *Two-Phase Commit Protocol for Mobile Wireless Environment (M-2PC)* [10] is to adapt the 2PC protocol for mobile systems with distributed transactions. MHs are *active* participants in the execution of a transaction and they send confirmation (that the work is done) to the agent or to the fixed device, in order to save energy. This model requires simultaneous connection of all mobile participants at the beginning of a transaction. The protocol does not provide adequate management of mobility and failures caused by network disconnection, nor does it provide a mechanism to control the competitiveness of distributed transactions.

The *Fault-Tolerant Pre-Phase Transaction Commit (FT-PPTC)* protocol [11] provides mechanisms for dealing with disturbances in the system in mobile environment, and supports heterogeneous mobile databases. FT-PPTC implements distributed transaction in two phases, i.e. *pre-phase*, one which is covering the MHs and the *main phase* which refers to the fixed part of the network. MHs are *active* participants in the execution of a transaction, but no mechanisms are developed for competition in mobile distributed transactions. FT-PPTC does not provide adequate management of mobility, as well, because when MHs are disconnected from the fixed network for a long time, they can block resources on the fixed participants. This, in turn, leads to an increased number of mobile transaction aborts.

In the *concurrency control without locking* approach [12], a concurrency control mechanism in mobile environment is proposed, by introducing the concept of *Absolute Validity Interval (AVI)* which is a time period in which the data item is said to be valid. The new mechanism provides reading the same available data item by multiple MHs – if the data item is updated by any mobile host, the MHs which have already read the same value must be invalidated. It reduces the waiting time for execution of a transaction and resources are not unnecessary locked.

## 3. MODEL OF THE MOBILE ENVIRONMENT

In this paper we consider a system model for the mobile distributed environment consisting of a set of MHs and a set of *fixed hosts (FHs)*, presented in Figure 1. The model has two main parts of the network: fixed and mobile. Communication between the two is conveyed via *Mobile Support Stations (MSS)*, which are connected to the fixed part of the network via wired links. MHs can cross the border between two different geographical areas covered by different MSSs.

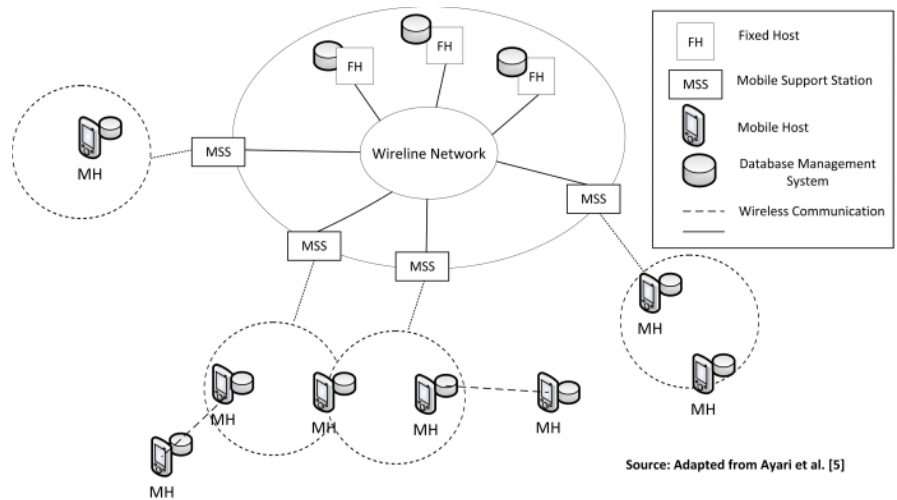

Figure 1. Communication model in the mobile environment

In the system model under consideration, MHs can communicate with the FHs through a MSS via wireless channels only when they are located within the MSS coverage area. Our *first improvement* is that MHs can *ad-hoc communicate* with neighboring MHs via wireless channels. When MHs enter a geographical area that is out of coverage of any MSS, in order to access database servers in the fixed network, they may connect through a neighboring MH which is in the covering area of any MSS. Thus, the nodes in a wireless network are responsible for both sending their traffic as well as relaying the traffic of other nodes in the network [21]. In mobile environments, the problem of appropriately decentralizing *replicated* servers has been investigated in [22], in order to enhance users' access speed in a wireless network.

In brief, we consider a mobile distributed environment where (i) MHs can communicate with each other (and/or with FHs) through MSS and, in addition, (ii) MHs can ad-hoc communicate with neighboring MHs in order to reach the fixed part of the network. We assume that database servers are installed on each FH, and each MH has a mobile database server installed.

## 4. THE TRANSACTION MODEL

A distributed transaction where *at least one* MH participates is called a *mobile transaction*. We identify a MH where a transaction is issued as a *Home-MH (H-MH)*. Participating MHs and FHs in the execution of a mobile transaction are called *participant MHs (Part-MH)* and *participant FHs (Part-FH)*, respectively.

We assume the existence of a *Coordinator (CO)*, which is responsible for coordinating the execution of the corresponding transaction. The CO is responsible for storing information concerning the state of the transaction execution. Based on the information collected from the participants of the transaction, the CO takes the decision to *commit* or *abort* the transaction and informs all the participants about its decision. The CO should be executed on a fixed host or hosts. This means that logs will be kept more safely.

## 5. CONNECTION FAULT-TOLERANT COMMIT PROCESSING FRAMEWORK

### 5.1. Overview

Some of the most frequent failures in mobile environments are communication failures. When MHs are in motion, they may exit the geographical area that is covered by some MSS and the resources of the fixed participants may potentially be blocked for an undefined period of time. If MHs do not reestablish connection with any MSS the transaction is aborted.

To minimize the number of mobile transaction *aborts* by tolerating failures caused by network disconnections and reduce the resource *blocking times* of fixed participants, we propose a CFT model for distributed transaction processing in mobile computing environment. The CFT framework ensures the atomicity property and its main objective is to help protocol designers and/or application designers in commit processing in mixed (infrastructure-based and ad-hoc supported) mobile distributed environment.

The CFT model involves two communication protocols and a decision algorithm, as well:

1. The first protocol is a *Standard communication protocol*, when MHs can *directly* connect to the fixed part of the network through MSSs;
2. The second one is an *Ad-hoc communication protocol*, when MHs cannot directly connect to the fixed part of the network through any MSS; with this, MHs are allowed to *ad-hoc communicate* with neighboring MHs in order to reach the fixed part of the network.

In the Standard communication protocol, similar as in [23], to minimize the use of wireless communication and to maximize conservation of MHs' resources, to each MH we assign a *Mobile Host Agent (MH-Ag)* that we add to the fixed network. We assume that in the execution of a transaction MH-Ag is representing the MH in the fixed network and it acts as an *intermediator* between the MH and the transaction CO, i.e. all the communication between MH and CO goes through the MH-Ag. The MH-Ag is responsible for storing all the information related to the state of all mobile transactions involving the MH. In the fixed network, a server (or servers) can be designated, where MH-Ag is created for each participating MH.

The second (ad-hoc) communication protocol comes into play when MHs cannot directly connect to the fixed network, or a MH cannot directly communicate with its MH-Ag through any MSS. In that case, MHs try to connect via ad-hoc communication with any neighboring MH which is in the covering area of any MSS. To allow this, we assign a *MH-Relay Agent (MH-RAg)* to each MH, which is responsible for ensuring a relay-wireless-link between neighboring MHs. In order to preserve generality, the CFT Model is built under the assumption that MH-RAg can use just about *any* of the well-known ad-hoc communication protocols.

Furthermore, as our *second improvement*, we define an additional function of a MH-Ag that we call a *Decision Algorithm (DAlg)*. DAlg is used during the execution of a transaction when MH-Ag cannot directly nor ad-hoc communicate with its MH for a predefined period of time. DAlg's task is to check if *Transaction Processing Fragment (TPF)* function is either **WRITE** (insert/update/delete) or **READ**. If the former is the case**,** DAlg saves the TPF in a FIFO (First-In First-Out) queue list and makes a decision for the MH to send "Yes" vote to the transaction CO. When the connection between MH and the corresponding MH-Ag is reestablished, MH-Ag's first task is to send all saved TPFs to the corresponding MH. Otherwise, if the TPF function is **READ**, DAlg makes a decision for the MH to send "No" vote to the transaction coordinator.

Yet again, the rationale behind the employment of DAlg is to help protocol and application designers to deal with commit processing in distributed mobile environment when neither standard nor ad-hoc communication is available. The number of methods (i.e. functions) that the DAlg implements can be freely specified by application designers, in order to reach their specific application needs.

### 5.2. Connection Fault-Tolerant Model Operation

In this section we illustrate the execution of a mobile transaction (Figure 2) in the proposed CFT commit processing framework.

If H-MH (Table 1) is connected to the fixed network through some MSS, it initiates a mobile transaction by sending TPF to the transaction CO (Table 2) through its corresponding MH-Ag (Table 3), which acts as an *intermediator* between CO and MH.

Transaction CO computes the *Execution timeout (Et)*, which is a time limit for all participants to complete the execution of the TPFs and send a vote to CO. After that, CO sends Et and TPFs to all Part-FHs and MH-Ags that represent the Part-MHs (Table 5) in the fixed network, asks them to PREPARE to commit the transaction, and enters the wait state. Subsequently, every MH-Ag initiates *Connection timeout (Ct)*, which is a time limit for MH-Ag to establish connection with its MH, and try to send Et and TPFs to its MH through either standard or ad-hoc communication protocol. If MH-Ag cannot establish connection with its MH before Ct expires, it activates the DAlg (Table 4).

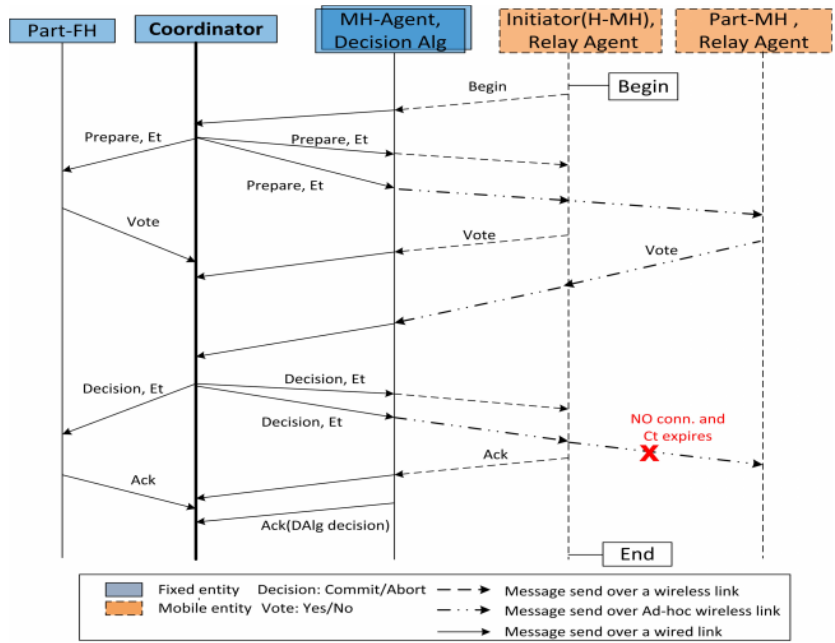

Figure 2. Execution of a mobile transaction in mobile environment

DAlg checks whether the *TPF* function is **WRITE** or **READ**. In the case of the former**,** DAlg saves the TPF in a FIFO queue list and makes a decision for the MH to send "Yes" VOTE to the transaction CO. As already mentioned before, if the TPF function is **READ**, DAlg makes a decision for the MH to send "No" VOTE to the transaction CO. If MH-Ag establishes connection with its MH before Ct expires, it sends TPF to its MH and resets Ct.

When the participants receive the PREPARE message, they check if they could commit the transaction. If so, and if MH establishes connection with MH-Ag before Ct expires, MH sends "Yes" VOTE to CO through its corresponding MH-Ag via standard or ad-hoc communication protocol. If Ct expires, MH-Ag activates DAlg.

Table 1. Home-MH's Algorithm

| Algorithm 1: H-MH's Algorithm |
|---|
| **1** Initialize Transaction (Ti); |
| **2 send** *TPF to its corresponding MH-Ag;* |
| **3 wait** for execution timeout (Et) and TPF; |
| **4** **if** *H-MH decides to abort Ti* **then** |
| **5**     **write** Abort record in the local log; |
| **6**     **If** H-MH is disconnected from the Fixed Network **then** |
| **7**       **send** Abort message to MH-RAg; |
| **8**     **else send** Abort message to its MH-Ag; |
| **9**     **return;** |
| **10 else** // H-MH decides to commit Ti |
| **11**     **force write** updates to the local log; |
| **12**     **If** H-MH is disconnected from the Fixed Network **then** |
| **13**       **send** updates to MH-RAg; |
| **14**     **else send** updates to its MH-Ag; |
| **15**     **wait for** decision message from CO; |
| **16**     *if decision message is* **Commit then** |
| **17**       **commit** *Ti;* |
| **18**       **write** Commit record in the local log; |
| **19**       **If** H-MH is disconnected from the Fixed Network **then** |
| **20**         **send** Ack message to MH-RAg; |
| **21**       **else send** Ack message to its MH-Ag; |
| **22**       **return;** |
| **23**     **else** //decision message is Abort |
| **24**       **abort** *Ti;* |
| **25**       **write** Abort record in the local log; |
| **26**       **If** H-MH is disconnected from the Fixed Network **then** |
| **27**         **send** Ack message to MH-RAg; |
| **28**       **else send** Ack message to its MH-Ag; |
| **29**       **return;** |
| **30**     **End** |
| **31 end** |

Table 2. Coordinator's Algorithm

| Algorithm 2: Coordinator's Algorithm |
|---|
| **1 wait for** Ti from H-MH; |
| **2 compute** Execution timeout (Et); |
| **3** extract TPFs of the Part-MHs/Part FHs and sends them with Et to MH-Ags and Part-FHs; |
| **4 while** *waiting for Et to expire* **do** |
| **5**     **if Abort** *message is received from the Part-MHs* **then** |
| **6**       **send Abort** to all FHs and all MHs through theirs MH-Ags; |
| **7**       **return;** |
| **8**     **End** |
| **9 end** |
| **10 if** *receive messages from each Part-MHs through MH-Ags and from FHs before Et expires* **then** |
| **11**     **write** all received updates; |
| **12**     *if all votes were Yes* **then** |
| **13**       **send Commit** to MH-Ags and to Part-FH; |
| **14**       **return;** |
| **15**     **else** //at least one of the votes is No |
| **16**       **send Abort** to MH-Ags and Part-FHs; |
| **17**       **return;** |
| **18**     **End** |
| **19 else** // Et expires |
| **20**     **send Abort** to MH-Ags of all Part-MHs; |
| **21**     **return;** |
| **22 end** |

After the CO has received VOTE from every participant, it decides whether to COMMIT or ABORT the transaction. If, for any reason, even *one* of the participants votes "No" or Et expires, the CO decides to ABORT the transaction and sends "Abort" message to all participants. Otherwise, if all the received votes are "Yes" and Et has not expired, the CO decides to COMMIT the transaction and sends "Commit" message, with reset Et to all participants. The participants need to ACKNOWLEDGE the CO's decision before the reset Et expires.

Table 3. MH-Ag's Algorithm

| **Algorithm 3: MH-Ag's Algorithm** |
|---|
| **1** **while** *working* **do** |
| **2**    **if** some MH reconnects to the fixed network **and** DAlg's FIFO queue is not empty **then** |
| **3**      **send** all stored TPF to its MH; |
| **4** **end** |
| **5** **wait for** receiving massages from CO or from its MH; |
| **6** **for** *any received message* **do** |
| **7**    **if** *message is sent by the CO* **then** |
| **8**      **compute** Connection timeout (Ct); |
| **9**      **if** MH-Ag is connected with MH through Standard or Ad-hoc conn. **and** Ct>0 **then** |
| **10**        **send** the received message to the corresponding Part-MH; |
| **11**        **reset** Ct; |
| **12**      **else** Execute Decision Algorithm; (**Algorithm 4**) |
| **13**    **else** //*message is send by corresponding Part-MH* |
| **14**      **if** MH is connected with MH-Ag through MSS or Ad-hoc conn. **and** Ct>0 **then** |
| **15**        **send** the received message to the CO; |
| **16**      **else** Execute Decision Algorithm; (**Algorithm 4**) |
| **17**    **end** |
| **18** **end** |

Table 4. Decision Algorithm           Table 5. Part-MH's Algorithm

| **Algorithm 4: Decision Algorithm** |
|---|
| **1 if** TPF is for Changing (Insert, Update, Delete) **then** |
| **2**    **save** TPF in FIFO queue; |
| **3**    **send** Yes(Presumed Commit) / ACKNOWLEDGE vote to CO; |
| **4 else** // *That means that TPF is for Reading* |
| **5**    **send** No (Presumed Abort) / ACKNOWLEDGE vote to CO; |
| **6 end** |

| **Algorithm 5: Part-MH's Algorithm** |
|---|
| **1 wait for** receiving the corresponding TPF and Et; |
| **2** //*Continue with step 4 to step 31 of Algorithm 1 substituting H-MH with Part-MH* |

## 6. DISCRETE-EVENT SIMULATION RESULTS AND ANALYSIS

In our simulation-based performance analysis of the Connection Fault-Tolerant Model, we focus on the *mobile transaction commit rate* performance metric. The analysis is conducted using SimPy [24] – an object-oriented, process-based, general-purpose discrete-event simulation package based on standard Python programming language [25]. A simulation run is set to simulate 10 hours. Transactions are generated with *exponentially distributed* interarrival times with a *mean* of 30 seconds, under the assumption that all transactions are of similar length, but experience different connection conditions. The number and the nature of Part-MHs and Part-FHs are randomly selected in order to model arbitrary heterogeneity. Table 6 summarizes our simulation parameters.

Table 6. Simulation settings

| Parameter | Value |
|---|---|
| Number of Part-MHs | 3 – 5 |
| Number of Part-FHs | 1 – 5 |
| Fragment execution time (Part-MH) | [0.3 – 0.7]s |
| Fragment execution time (Part-FH) | [0.1 – 0.3]s |
| Transmission delay (wireless link) | [0.2 – 1.0]s |
| Transmission delay (wireless ad-hoc link) | [0.4 – 2.0]s |
| Transmission delay (wired link) | [0.01 – 0.03]s |
| Disconnection Rate | [0 – 95]% |
| Ad-hoc support | [10 – 90]% |
| Distributed transaction WRITE function | [10 – 90]% |

The *disconnection rate* is defined as the ratio of time when the participating MH is disconnected from the fixed network, against the total simulation time. Similarly, the *level of ad-hoc support* is the ratio of the time when ad-hoc communication is available between MHs, against the total simulation time. It is hard to quantify the level of ad-hoc support between MHs in mobile distributed environment, i.e. in some parts of the wireless network it can be significantly lower compared to other. Consequently, we distinguish between *three* groups that represent different parts of the wireless network with different levels of ad-hoc support (e.g. there is a known fundamental relationship between *node density* and *delay* in wireless ad-hoc networks with *unreliable* links). Every MH in the wireless network is a member of one of the predefined groups.

Considering several different scenarios *with-* or *without* support of ad-hoc communication and *with-* or *without* support of DAlg, our simulation-based performance analysis aims to evaluate the performance potential of the Connection Fault-Tolerant commit processing framework, as well as to assess its contribution to the overall mobile transaction commit rate.

### 6.1. Ad-Hoc Communication Impact on Transaction Commit Rate

Simulation experiments study the impact of ad-hoc communication on mobile transactions commit rate. Figures 3-5 illustrate the mobile transaction commit rate against different disconnection rates, and different values representing different levels of ad-hoc support. Arbitrarily, transaction timeout is set to 5 seconds.

The results shown in Figure 3 resemble the case where all MHs are members of a single group, i.e. the level of ad-hoc support is the same in any part of the wireless network. One can conclude that the ad-hoc support in the CFT model considerably improves the transaction commit rate, and the ad-hoc communication impact is higher for networks where, actually, the disconnection rate is higher.

Figure 4 shows simulation results where MHs are classified in *two* different groups that have different levels of ad-hoc support. It is evident that ad-hoc support improves the transaction commit rate, but the percentage of improvement is lower compared to the previous scenario, where the level of ad-hoc support was the same in each part of the wireless network. In addition, commit rate slightly decreases when the disconnection rate rises all the way up from 60 to 95%.

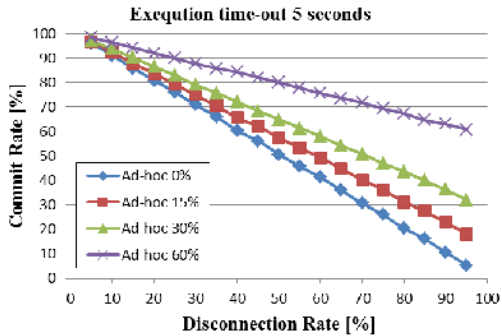
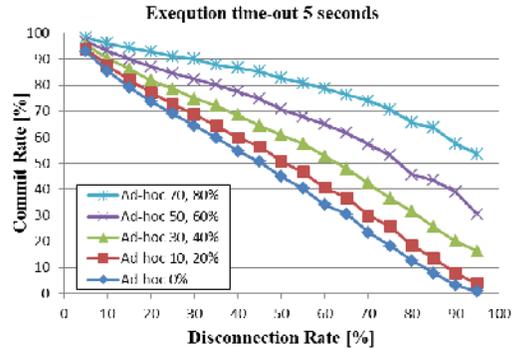

Figure 3: Impact of ad-hoc communication on commit rate (single group)

Figure 4: Impact of ad-hoc communication on commit rate (two groups)

At the end, in order to present the influence of ad-hoc support in a *highly dynamic* wireless network, we classify all MHs in *three* groups that have different levels of ad-hoc support. From the chart in Figure 5, one can conclude that commit rate increment is not evident as before, i.e. the highest improvement of the commit rate is only about 11%.

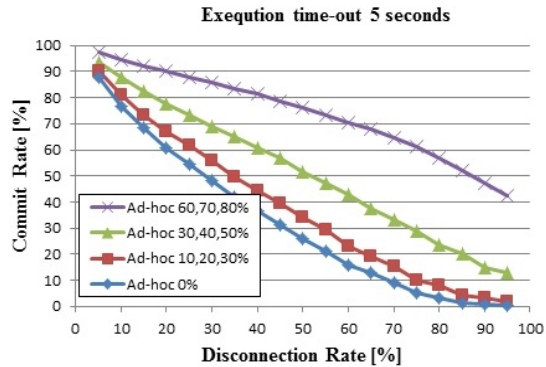

Figure 5: Impact of ad-hoc communication on commit rate (three groups)

### 6.2. The Rationale behind the Employment of a Decision Algorithm

In respect of the stochastic nature of random network disconnections, we need to determine *Connection timeout* values that contribute the most to the highest perceived ad-hoc communication impact, *before* activating the DAlg. Figure 6 shows the mobile transaction commit rate against different connection timeouts and different ad-hoc support values. Execution timeout is set to UNLIMITED. We assume that the functions of all mobile transactions are READ, meaning that if DAlg activates, the transaction will be aborted. From the chart in Figure 6, one can conclude that for any level of ad-hoc support, commit rate increment is more evident for increments of connection timeout up to 2.4 seconds – afterwards, the commit rate only slightly increases with the rise of the connection timeout time. This induces that connection timeout value of 2.4 seconds is the *optimum value* for maximum ad-hoc communication contribution to the mobile transaction commit rate in the CFT model.

To determine the combined ad-hoc communication and DAlg contribution to the transaction commit rate, we performed numerous simulations for *sparse* (Figures 7, 8) and *dense* environments (Figures 9, 10).

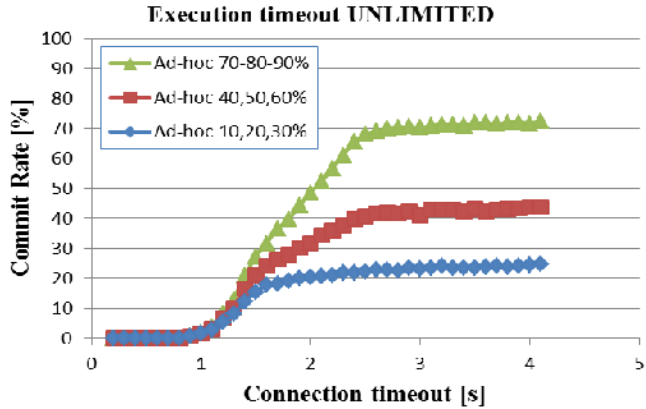

Figure 6**.** Impact of Connection timeout on commit rate

Figures 7 and 9 show the mobile transaction commit rate against different disconnection rates and different percentages of mobile transactions whose function is WRITE. Connection timeout is set to the optimum value of 2.4 seconds. The level of wireless ad-hoc support for the Figure 7 is set to 10, 20 and 30% for the *three groups* of MHs, whereas in Figure 9 the level of ad-hoc support varies between 70 and 90%. The first curve ("standard") in Figures 7 and 9 presents results when only standard communication is available between MHs and the fixed network. In other words, we have an execution of the standard 2PC protocol. The second curve ("ad-hoc") presents results when, besides the standard communication, ad-hoc communication between MHs is available, as well. Compared to the first scenario, these results show slight increment of the transaction commit rate. The rest of the curves present results where, besides standard and ad-hoc communication, DAlg is employed. It is evident that (i) the mobile transaction commit rate is higher if the percentage of mobile transactions, whose function is WRITE, is higher and (ii) a higher level of ad-hoc support leads to a higher commit rate. Moreover, it can be concluded that both ad-hoc communication and the employment of DAlg increase the transaction commit rate (compared to 2PC).

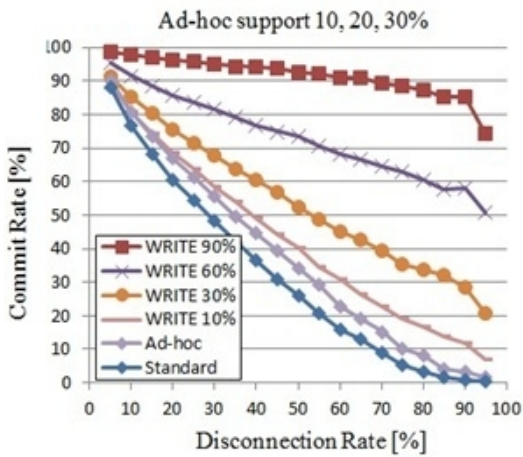 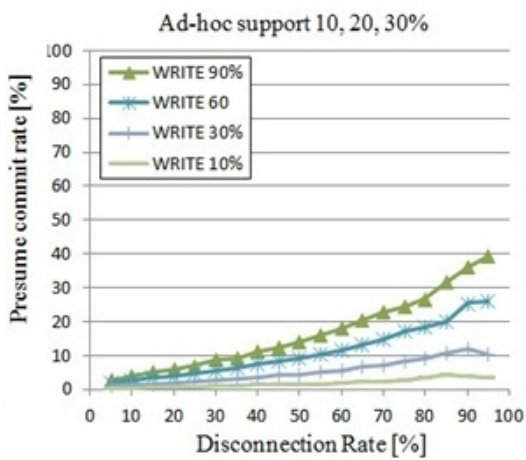

Figure 7. CFT contribution to commit rate    Figure 8. DAlg contribution to commit rate

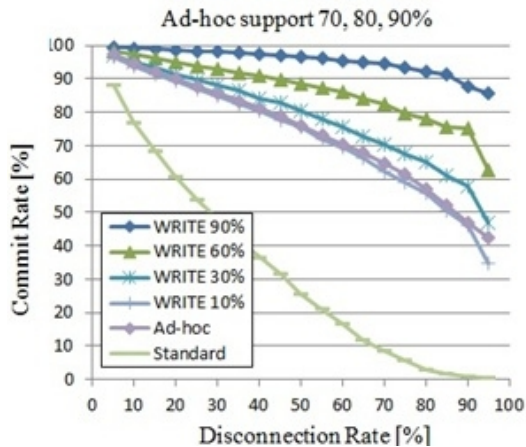 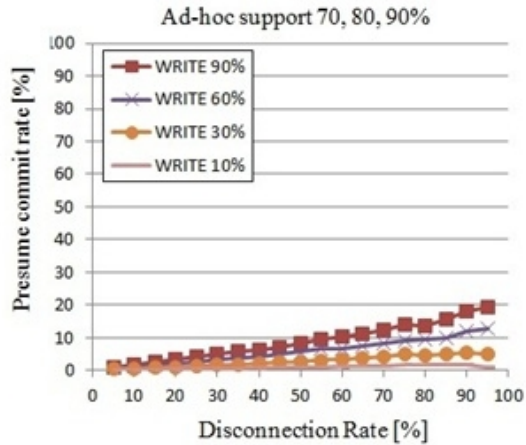

Figure 9. CFT contribution to commit rate    Figure 10. DAlg contribution to commit rate

Figures 8 and 10 show mobile transaction *presumed-commit rate* against different disconnection rates and different percentages of mobile transactions whose function is WRITE. Presumed-commit rate is the percentage of committed mobile transactions where DAlg was actively involved. It is evident that DAlg contribution on transaction commit rate is higher when disconnection rate rises.

From the charts, one can conclude that if the level of ad-hoc support is low, the impact of DAlg on the mobile transaction commit rate is rather strong (Figure 8). The other way around, when the level of ad-hoc support is high, the impact of DAlg on the mobile transaction commit rate is diminishing (Figure 10).

## 7. CONCLUSION

In this paper we made a detailed review of the operation of a Connection Fault-Tolerant Model for distributed mobile transaction processing. In order to preserve generality, the commit processing framework is built under the assumption that mobile hosts can use just about *any* of the well-known ad-hoc communication protocols, and the rationale behind the employment of a decision algorithm is to help protocol and application designers to deal with commit processing in distributed mobile environment when neither standard nor ad-hoc communication is available.

As a general conclusion from our simulation-based performance analysis, the *partial contribution* of (i) ad-hoc communication between mobile hosts and (ii) the employment of an appropriate decision algorithm for mobile host agents is *counter proportional*: when the level of ad-hoc support is lower, the impact of the employment of a decision algorithm on the mobile transaction commit rate is stronger and vice versa. Nevertheless, conjointly, they substantially improve transaction commit rate when compared to the standard Two-Phase Commit (2PC) protocol. Moreover, there is a threshold effect on the connection timeout values: *below* the threshold the ad-hoc communication contribution *increases* with the timeout value, whereas *above* the threshold the contribution is close to a limit – *additional* increase in the overall mobile transaction commit rate can be seen only with the employment of a decision algorithm.

In our future work we plan to expand the parameters that the decision algorithm uses for decision making on behalf of a MH, as well as to employ the class of Deterministic and Stochastic Petri Nets (DSPNs) for modeling and analysis of the Connection Fault-Tolerant commit processing framework (e.g. *deterministic* transitions – to capture the timeout mechanisms, and *exponential* transitions – to capture the interarrival times of transactions, as well as the stochastic nature of random network disconnections), in order to evaluate a range of *availability*, *reliability*, *performance* and *performability* measures.

**Authors**

**Tome Dimovski** received his BSc and MSc degrees from the Technical University – Gabrovo, Bulgaria, in 2004 and 2005, respectively. He is a teaching and research assistant at the Faculty of Technical Sciences, St. Clement Ohridski University in Bitola, Republic of Macedonia, and currently works on his PhD thesis at the Department of Computer Science and Engineering. His research interests include Distributed Databases, Computer Networks and Mobile Computing. He is a member of the ACM.

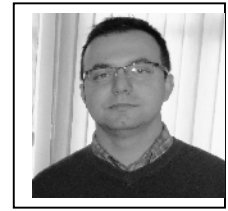

**Pece Mitrevski** received his BSc and MSc degrees in Electrical Engineering and Computer Science, and the PhD degree in Computer Science from the Ss. Cyril and Methodius University in Skopje, Republic of Macedonia. He is currently a full professor and Head of the Department of Computer Science and Engineering at the Faculty of Technical Sciences, St. Clement Ohridski University, Bitola, Republic of Macedonia. His research interests include Computer Networks, Computer Architecture, Performance and Reliability Analysis of Computer Systems and Stochastic Petri Nets. He has published more than 60 papers in journals and refereed conference proceedings and lectured extensively on these topics. He is a member of the IEEE Computer Society and the ACM.

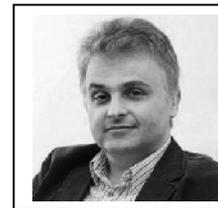